\documentclass[%
 reprint,
 amsmath,amssymb,
 aps,
]{revtex4-2}
\usepackage{float}
\usepackage{graphicx}
\usepackage{caption}
\usepackage{subcaption}% Include figure files
\usepackage{dcolumn}% Align table columns on decimal point
\usepackage{bm}% bold math
\usepackage{tabularx}
\usepackage{amsmath}
\usepackage[T1]{fontenc}

\begin{document}

\preprint{APS/123-QED}

\title{Quantum pseudo-integrable Hamiltonian impact systems.}% Force line breaks with \\

\author{Omer Yaniv and Vered Rom-Kedar}
 
 \affiliation{Department of Computer Science and Applied Mathematics,
Weizmann Institute of Science, Rehovot 7610001, Israel}%Lines break automatically or can be forced with \\

\date{\today}% It is always \today, today,

\begin{abstract}
Quantization of a toy model of a pseudointegrable Hamiltonian impact system is introduced, including EBK quantization conditions, a verification of Weyl's law,  the study of their wavefunctions and a study of their energy levels properties. It is demonstrated that the energy levels statistics are similar to those of pseudointegrable billiards. Yet, here, the density of wavefunctions which concentrate on projections of classical level sets to the configuration space does not disappear at large energies, suggesting that there is no equidistribution in the configuration space in the large energy limit; this is shown analytically for some limit symmetric cases and is demonstrated numerically for some nonsymmetric cases.
\end{abstract}

\maketitle

%\tableofcontents

Quantum chaos studies how classical dynamics (integrable and non-integrable) are reflected in the properties (e.g. eigenvalues and eigenfunctions) of the correspondent quantum system.
It is accepted that in integrable systems, the distribution of the level spacing is provided by the Poisson distribution $e^{-s}$ \cite{berry1977level}, while that in chaotic systems (hereafter, meaning mixing system on energy surfaces, studied by simulating chaotic billiards) they distribute as eigenvalues of random matrix ensembles (GOE)  \cite{bohigas1984characterization}. When a system has a mixed phase space, which is the common behavior of smooth Hamiltonian systems, it is found that a Berry-Robink distribution, a convex hall of the Poisson and the GOE distributions, describes the level spacing \cite{berry1984semiclassical,prosen1994numerical}. This distribution reflects the existence of eigenfunctions supported on the  islands of stability  and of eigenfunctions supported on the chaotic components of the classical phase-space \cite{backer2005flooding}.

Pseudointegrable dynamics, correspond to systems with intermediate complexity: the phase space trajectories are not ergodic on the full energy surface, yet, they are not always periodic or quasi-periodic. Such systems arise in the study of plane polygonal rational billiards (polygonal tables with all corners being rational fractions of $\pi$), where trajectories  move on invariant two-dimensional surfaces of genus $g > 1$ \cite{richens1981pseudointegrable,gutkin1996geometry}. The level spacing in such quantum systems appears to have intermediate statistics: the nearest-neighbor distribution displays repulsion at small distances and an exponential decay at large distances  \cite{bogomolny1999models}. 

Another important characteristic of quantum systems is the asymptotic distribution of  their wavefunctions. For systems with classical ergodic dynamics, in the semi-classical limit, the  eigenfunctions which are equidistributed form  a density 1 sequence \cite{zelditch1996ergodicity}.
In particular, such wavefunctions are equidistributed in both configuration space and momenta space. The other wavefunctions, which are not equidistributed, have scars - they concentrate along invariant phase space sets or on singular sets of the classical dynamics \cite{heller1984bound,hassell2010ergodic}. For chaotic billiards, the most visible scars are associated  with low period unstable periodic orbits and orbits at corners of the billiard table \cite{heller1984bound,cvitanovic2005chaos}.

Since plane rational polygonal billiards are ergodic only in the configuration space (and not in the momenta space), equidistribution of the wavefunctions can be expected only in their configuration representation. Following  \cite{zelditch1996ergodicity}, it was established that also here, in the semi-classical limit, scars in configuration space can only appear for a vanishing density of eigenfunctions  \cite{marklof2012almost}. Yet, it was observed, for finite energies, that some of the exceptional wavefunctions here have superscars; these concentrate on invariant sets associated with families of classical periodic orbits \cite{bogomolny2004structure}. 
Such structures were observed experimentally \cite{kudrolli1997experiments,bogomolny2006first}. 

In this letter we investigate eigenvalues statistics and eigenfunctions properties of a class of systems that belongs to the recently discovered family  of classical pseudointegrable Hamiltonian systems with impacts. Such systems combine motion under a smooth potential field with continuous symmetries and reflections from a corresponding family of billiards that keeps the continuous symmetries only locally and not globally. For example,  trajectories of a separable Hamiltonian  \begin{equation}
 H=H_1 + H_2, \  H_i(q_i,p_i)=\dfrac{p_i^2}{2m}+V_{i}(q_i), \ i=1,2 \label{eq:modelsham}\end{equation}
 in a right-angled polygonal billiard with at least one concave corner are pseudointegrable  \cite{becker2020impact,frkaczek2021non}. 
 
\begin{figure}[ht]
\begin{subfigure}{.2\textwidth}
 % \centering
  % include first image
  \includegraphics[width=.9\linewidth]{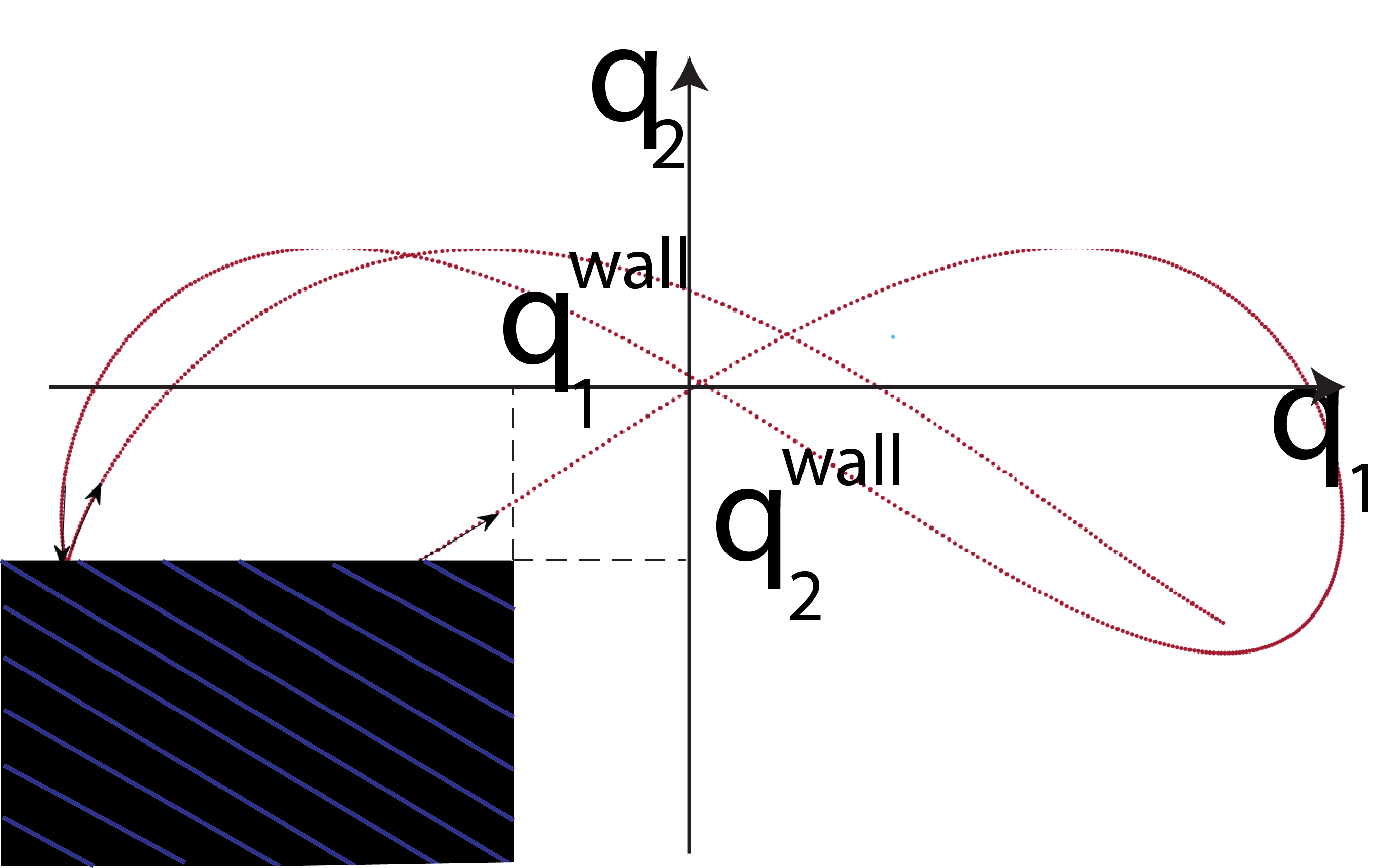}  
  \caption{}
  \label{fig:sub-first}
\end{subfigure}
\begin{subfigure}{.2\textwidth}
  %\centering
  % include second image
  \includegraphics[width=1.4\linewidth]{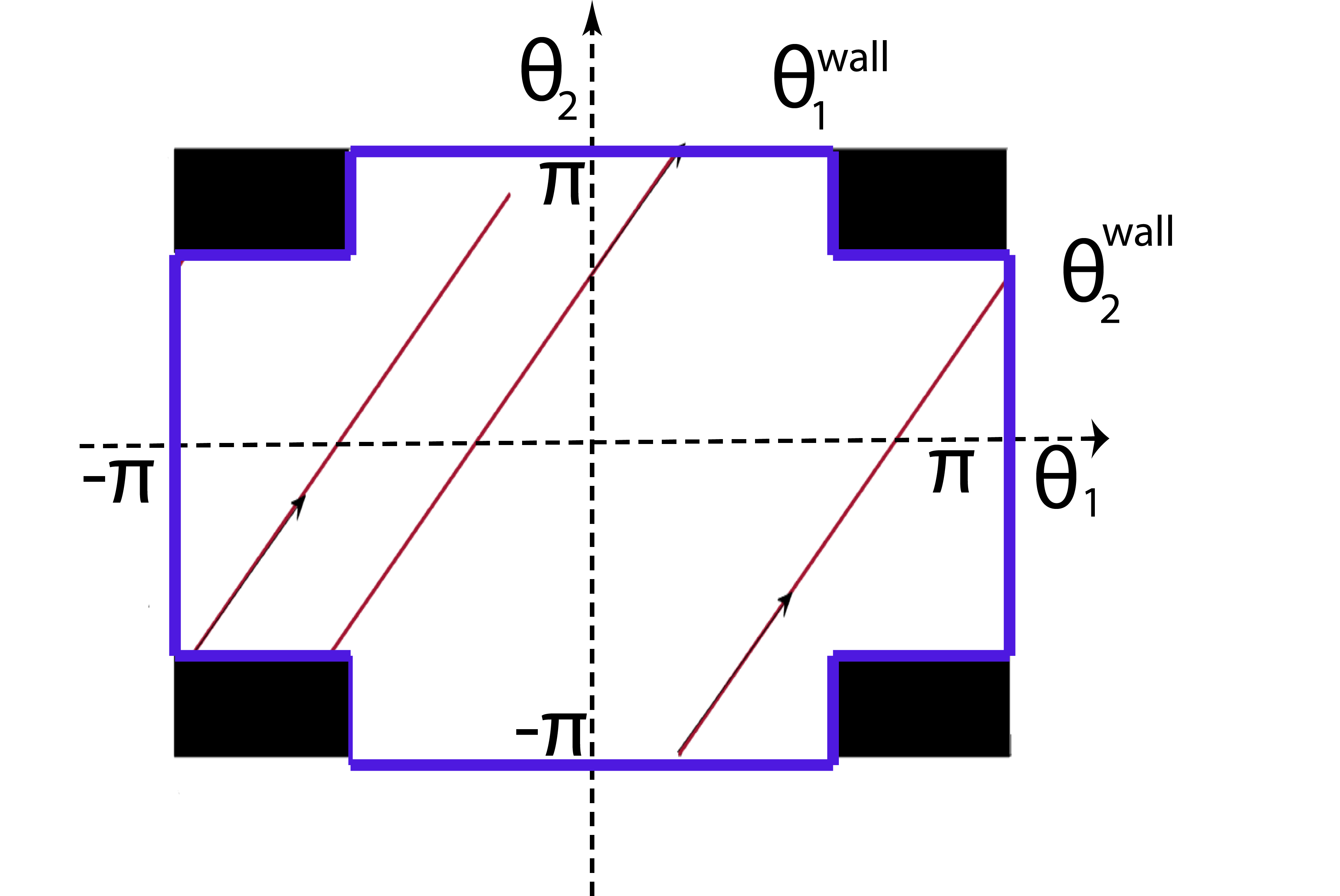}  
  \caption{ }
  \label{fig:sub-second}
\end{subfigure}
\begin{subfigure}{.2\textwidth}
%  \centering
  % include second image
  \includegraphics[width=.9\linewidth]{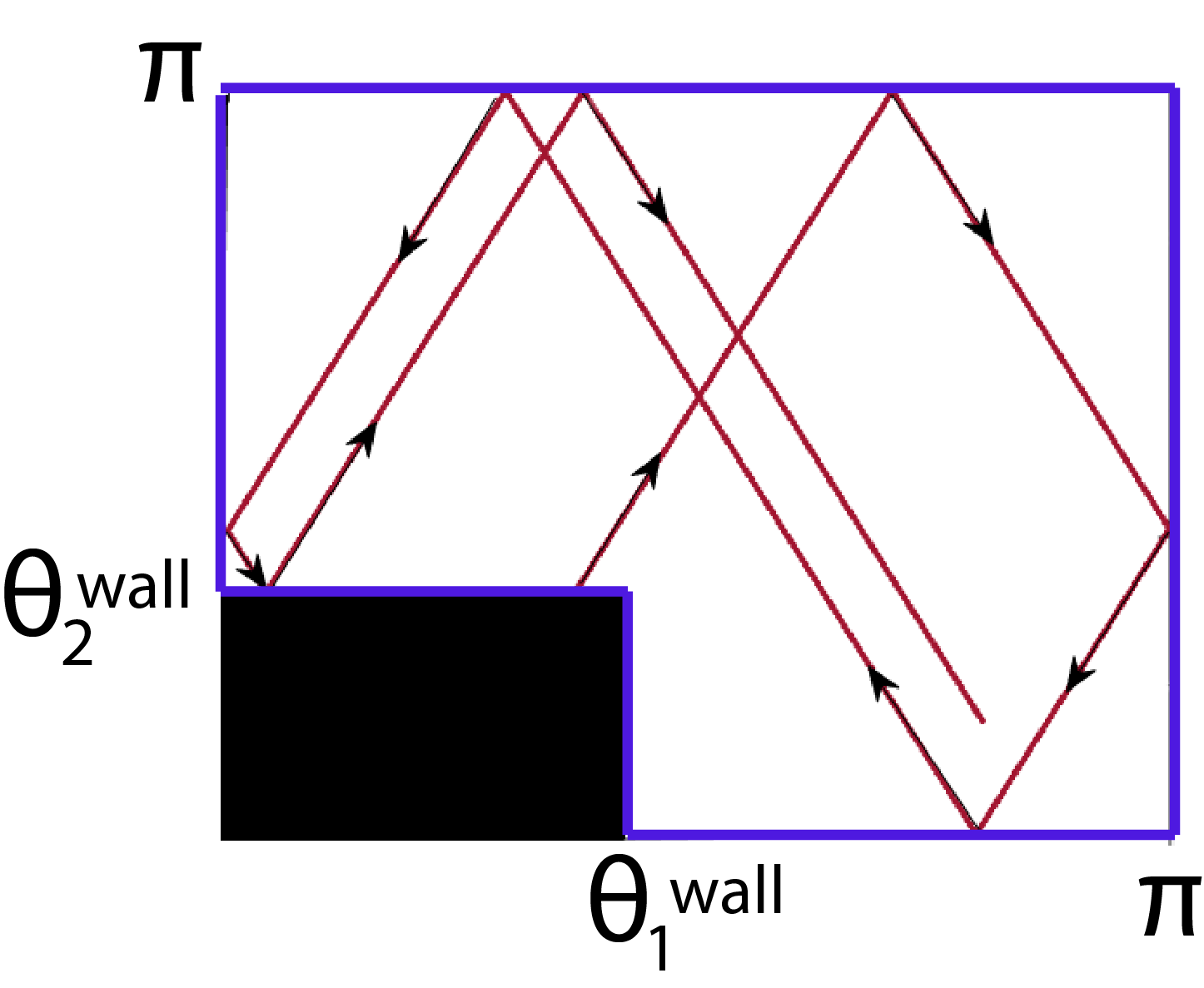}  
  \caption{}
  \label{fig:sub-second}
\end{subfigure}
\caption{A trajectory of a separable Hamiltonian  reflecting from a step.  (a)  Projection to the configuration space. (b)  The corresponding directed motion on the cross-shaped surface in the angles space. (c)  Folding the surface to the lower left quadrant leads to the corresponding billiard motion on an L-shaped billiard. Here, Eq. (\ref{eq:modelsham}) are integrated with elastic reflections from the step of Eq. (\ref{eq:stepdef}), with $V_i(q_i)=\frac{1}{2}\omega_i q_i^2, \omega_1=1,\omega_2=\sqrt{2}, q_1^{wall}=q_2^{wall}=-1, E_1=5.625,E_2=5.50 $.}
\label{fig:fig}
\end{figure}

 Here, we study the quantum step oscillators: we take $V_i$ to be confining potentials which are even smooth functions with a single minimum at the origin and are monotone elsewhere, and take the right angled polygon to be   \(\mathbb{R}^{2}\setminus S\), where \begin{equation}\label{eq:stepdef}
S_{q^{wall}}=\{(q_{1},q_{2})| \:q_1 < q_1^{wall}\le 0 \text{ and }  q_2 < q_2^{wall}\le 0\}. 
\end{equation}   The trajectories are confined by the potential and reflect from the step $S_{q^{wall}}$  \cite{becker2020impact}, see Figure  \ref{fig:fig}a. Since the step boundaries are parallel to the axes, the vertical and horizontal momenta are conserved at reflections, so the motion occurs along the level sets $H_i(q_i,p_i)=E_{i}, \ i=1,2$.   Passing to the action angel coordinates of the smooth separable system, provided $E_i>V_i(q_i^{wall}),\ i=1,2$, the motion on each level set is conjugated to the directed motion on the flat cross-shaped surface,  see  Figure \ref{fig:fig}b. The direction of motion on this surface is given by $\omega_2(E_2)/\omega_1(E_1)$ and the cross shaped concave corners are at  $\{\pm\theta_1^{wall}(E_1), \pm\theta_2^{wall}(E_2)\}$, where $\omega_i(E_i)$ denotes the frequency of the smooth periodic motion under $H_i$ and $\theta_i^{wall}(E_i)$ denotes the angle of an impacting trajectory (with the convention that $\theta_i=0$ at the maximum of $q_i$).    So, the direction of motion and the surface dimensions  depend continuously on $(E_1,E_2)$.
For the case of harmonic oscillators, i.e. when $V_i(q_i)=\frac{1}{2}\omega_i q_i^2$, the frequencies are fixed at $\omega_i $ and the values of $\theta_i^{wall}(E_i)$ can be explicitly computed.
Equivalently, by folding the surface, the motion on such level sets is conjugated to the directed billiard motion on an L-shaped billiard,  see Figure (\ref{fig:fig})c.  Thus,  this system is pseudointegrable \cite{becker2020impact}.
  In general, the dynamics on such surfaces has non-trivial ergodic properties. It was proven that if $q_i^{wall}<0$ for $i=1,2$, the motion is typically uniquely ergodic, and, for the case of resonant harmonic oscillators, there are level sets with co-existing periodic ribbons and dense orbits on some parts of the cross-shaped surface \cite{frkaczek2021non}.

As we are interested in quantization, and, in particular, in studying the role of superscars in the system, we look first for families of periodic orbits. Given a family of periodic orbits on a given level set $(E_1,E_2=E-E_1)$, with $\mu=(\mu_1,\mu_2)$ turning points ($\mu_1$ in the horizontal direction and $\mu_2$ in the vertical one), and $b=(b_1,b_2)$ impacts ($b_1$ with the right side of the step and $b_2$ with the upper part of the step), and an action $I(E;\mu,b)$, we can quantize it by using the EBK quantization conditions \cite{brack2018semiclassical,keller1958corrected}:
\begin{align}
\label{eqn:ebk}
I(E;\mu,b)=\hbar(n+\frac{\mu_1+\mu_2}{4}+\frac{b_1+b_2}{2}).
\end{align}
Moreover, denoting by $I_i(E_i)$ the action of the smooth $H_i$ system and by  $I_i^{wall}(E_i)=\int_{q_i\geq q_i^{wall}} p_i(q_i;E_i)dq_i=I_i \frac{2\theta_i^{wall}}{2\pi}$ the action of the impact $H_i$ system,  we obtain:
\begin{align}
\label{eqn:action}
I(E_1,E_2;\mu,b)=\sum_{i=1}^2 b_i I_i^{wall} + (\frac{\mu_i-b_i}{2})I_i,
\end{align}
namely, given  $\mu$, $b,I_i(E_i)$ and $ \theta_i^{wall}(E_i)$, we expect that the EBK quantization rule will predict the energy levels.  Yet, in general, it is non-trivial to find  $\mu$ and $b$ (see e.g. section 7 in \cite{frkaczek2021non}) nor to invert  $I(E_1,E_2;\mu,b)$ on the given family of periodic orbits.

\begin{figure}[htbp]
 \centering
 \begin{subfigure}[t]{.2\textwidth}
  % include second image
  \includegraphics[width=1\linewidth]{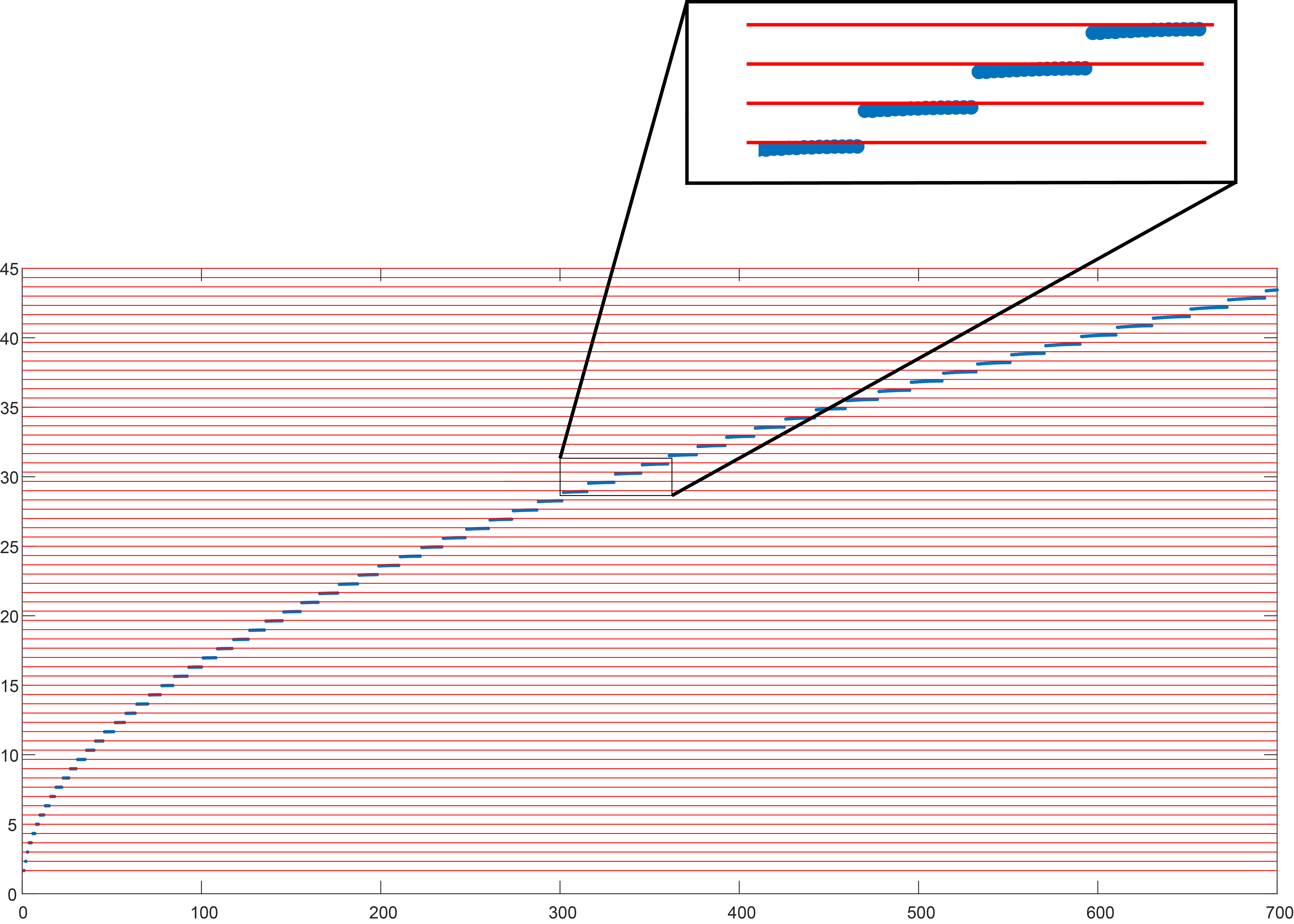}  
  \caption{}
  \label{fig:sub-second}
\end{subfigure}
\begin{subfigure}[t]{.2\textwidth}
  % include second image
  \includegraphics[width=1\linewidth]{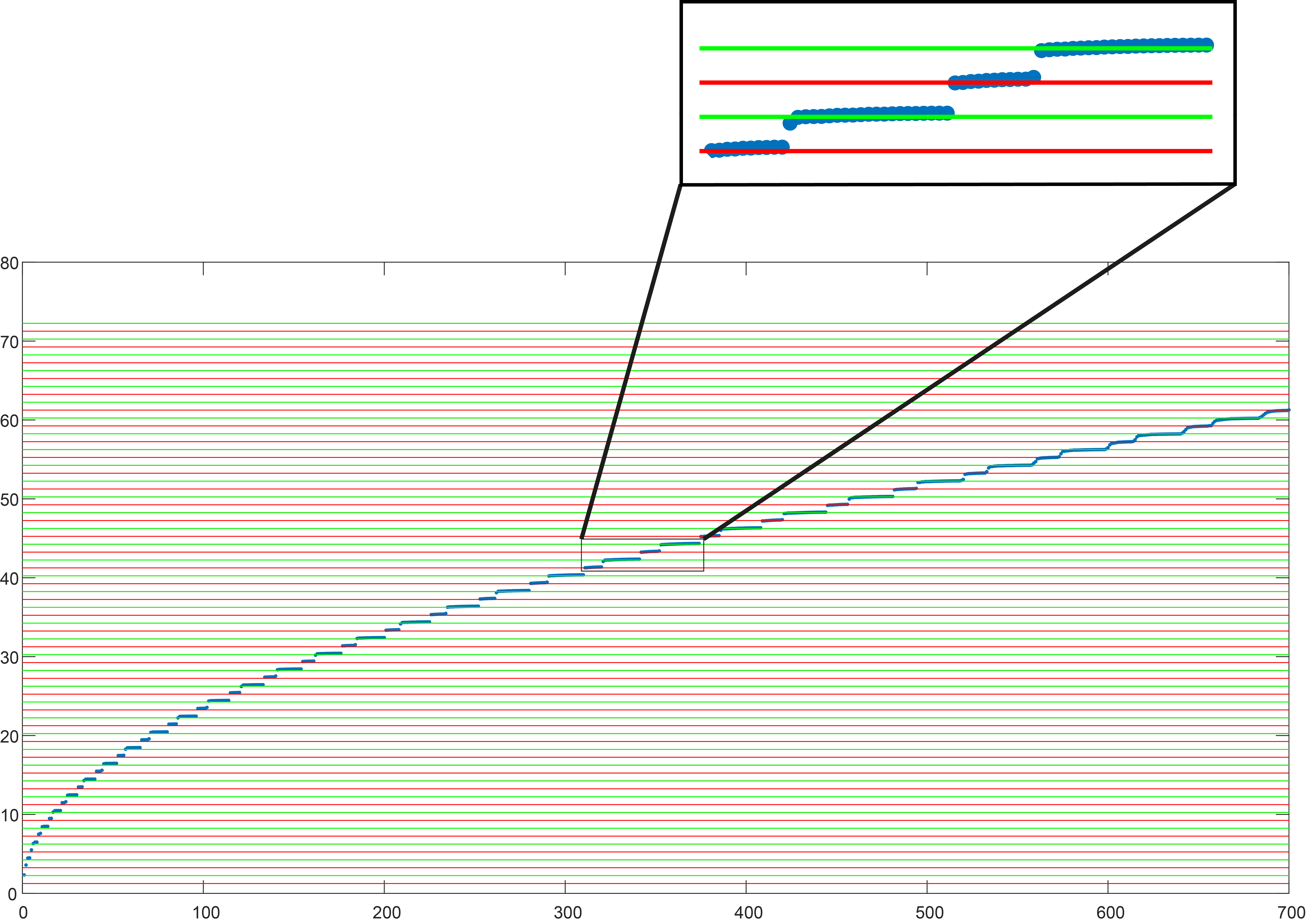}  
  \caption{}
  \label{fig:sub-second}
\end{subfigure}
    \caption{Energy levels for resonant harmonic oscillator with a step at the origin: numerical and expected (EBK) values. (a) Odd $m$ ($\omega_1=1,\ \omega_2=1$): the expected values for the single family of the periodic orbits of Eq. \ref{oddm} denoted by horizontal red lines agree with the numerical values (blue dots). (b) Even $m$ $\omega_1=1,\ \omega_2=2$: The expected values (family I:red and green horizontal lines, Family II: only green horizontal lines) agree with the numerical values, and the common values have larger degeneracy.}
      \label{figtable}
      \end{figure}

 We consider first some simple limit cases in which periodic motion can be easily identified. When the step is at the origin ($S_0=S_{q_1^{wall}=q_2^{wall}=0}$), the corner angles are fixed at $\theta_i^{wall}(E_i)|_{q_1^{wall}=q_2^{wall}=0}=\frac{\pi}{2}$, so the dimensions of the cross-shaped surface are independent of the energy. When the potentials are harmonic, the direction of motion, $\omega_{2}/\omega_{1}$ is independent of the energy as well and $I_i=E_i/\omega_i$. Thus, by choosing resonant harmonic potentials and a step at the origin, we conclude that for all partial energies the motion is periodic and of the same type and that $I_i^{wall}=I_i/2$. In particular, setting : $\omega_{1}=1, \omega_{2}=\frac{n}{m}$ (with $gcd(n,m)=1$), it can be shown that there are exactly  2 options for dynamics; When $m$ is odd there is  a single family of periodic orbits, whereas an even  $m$ leads to 2 distinct families of periodic orbits. In this latter case, one of the families has half of the action of the other one. Taking the simplest case of  $n=1$, we can compute the number of impacts and turning points for each of these families, and then, using Eqs. (\ref{eqn:ebk}) and (\ref{eqn:action})  provide  a prediction for the eigenvalues, $E_k$.
For odd $m$, we obtain that the periodic trajectory has $3(m+1)$ turning points ($\mu_1=3m,\mu_2=3$) and $m+1$ impacts ($b_1=m,b_2=1$), hence 
\begin{align} 
E_{k}=\frac{k}{1.5m}+\frac{5(1+m)}{6m}.\label{oddm}
\end{align}
For even $m$ we obtain that the first family of periodic orbits has  $2(m+1)$ turning points ($\mu_1=2m,\mu_2=2$) and $m$ ($b_1=m,b_2=0$) impacts, whereas the second one has $m+1$ ($\mu_1=m,\mu_2=1$) turning points and $1$ impact ($b_1=0,b_2=1$), hence 
\begin{align}
E^{I}_{k_1}=\frac{k_1}{m}+\frac{4m+2}{4m} \label{evenmfam1}\\ 
E^{II}_{k_2}=\frac{2k_2}{m}+\frac{m+3}{2m}\label{evenmfam2},
\end{align}
In figure \ref{figtable}, we validate the above results. Notice that for even $m$ there are infinite number of energy levels at which $E^{I}_{k_1}=E^{II}_{k_2}$ (marked with green lines), and in particular, for $m=2$, $E^{I}_{k_1}=E^{II}_{2k_1}$ (as shown in Fig.  \ref{figtable}).  Since the system here is symmetric, all these energy levels are degenerate, and, as shown in \ref{figtable}b, the common energy levels for the two families have higher degeneracy.
 
\begin{figure}[htbp].
  \centering
  % include first image
  \includegraphics[width=.9\linewidth]{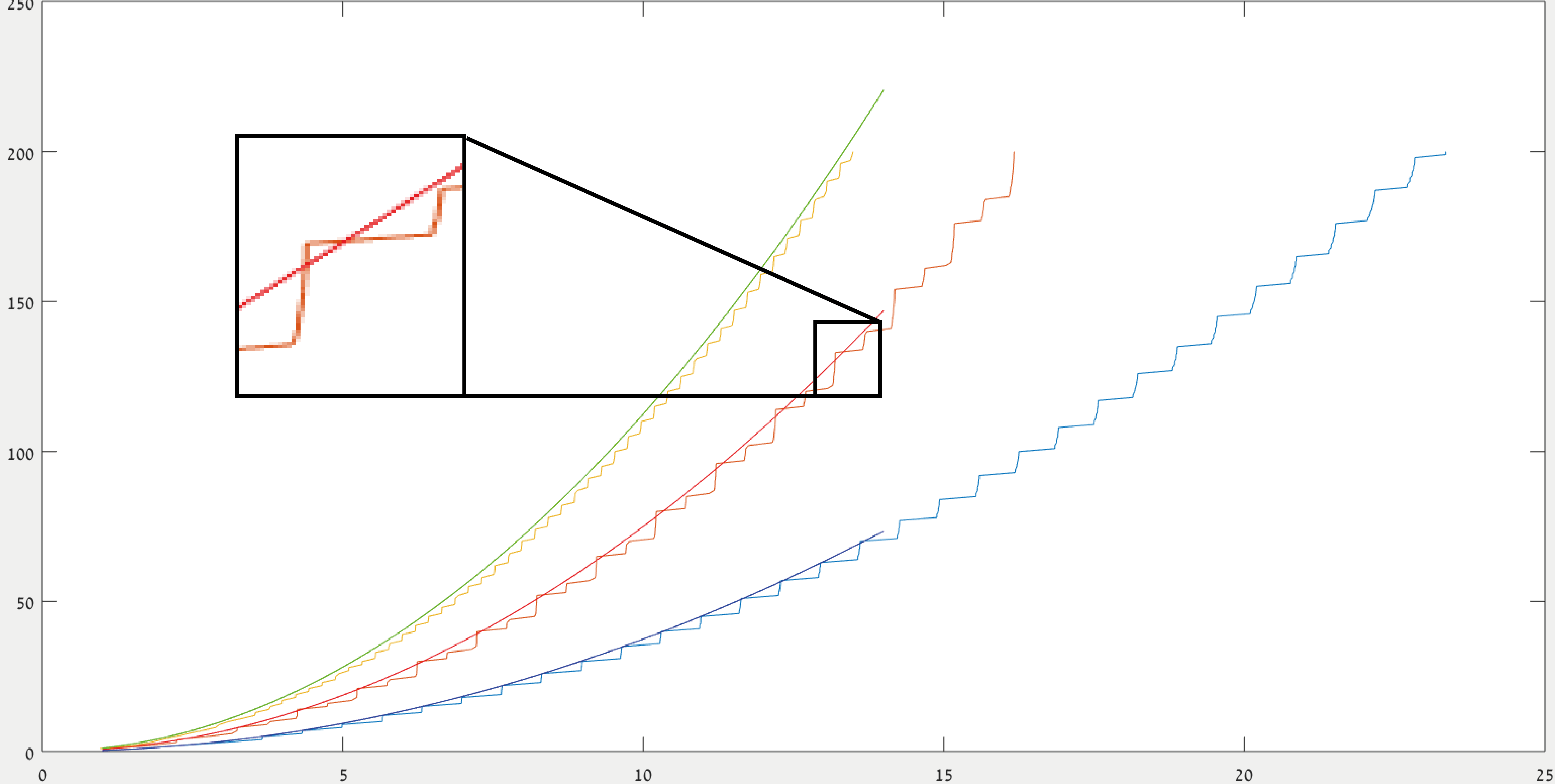}  
  \caption{Weyl's law. Smooth curves correspond to the predicted phase space volume (Eq.\ref{resvol}) for the three resonant cases ($\omega_1 =1,\omega_2=1,2,3$ yellow, red, blue lines respectively). These prediction fit the corresponding numerical results. The inset shows the non-uniform jump in $N$ for the even $m$ case.}
\label{fig2}
\end{figure}

Next we use Weyl's law to validate our computations of correspondence between the classical families of periodic orbits and the energy levels. Recall that for the two dimensional case, Weyl's law is:
\begin{align}
 N[E_{j}: E_{j} \leq b]=
\frac{1}{\hbar^2} \text{Vol}( H\leq b)+o(1)
\text{ as } \hbar\rightarrow 0 
\end{align}
and notice that the phase space volume for the step-oscillator is:
\begin{equation}
\begin{aligned}
\text{Vol}(\mathcal{E})=\int_{0}^{I_{2}(\mathcal{E})} dI_{2}\int_0^{I_{1}(\mathcal{E}-E(I_{2})))}-4\theta_1^{wall}(I_1)\theta_2^{wall}(I_2)\\+4\pi(\theta_1^{wall}(I_1)+\theta_2^{wall}(I_2))dI_{1}.
\end{aligned}
\end{equation}
For the case of a step at the origin and harmonic oscillators, we obtain \begin{equation}
\begin{aligned}
\text{Vol}(\mathcal{E})|_{S_0,\text{Harmonic oscillators}}=\frac{3\pi^2}{2\omega_{1}\omega_{2}}\mathcal{E}^2.
\end{aligned}
\label{resvol}
\end{equation}  Fig. \ref{fig2} shows this expected correspondence.
 For the even $m$ case the contribution of the larger degeneracy associated with the energy levels which are common to the 2 different families is evident.

Next, we examine non-resonant oscillators (and not necessarily harmonic) while keeping the step at the origin.  Classically, the motion is ergodic within the level set for almost all partial energies.
Hence, we expect wavefunctions to concentrate on the projection of such level sets to the configuration space. We show that at least for a sequence of density $\frac{1}{3}$ of the wavefunctions this property holds and doesn't vanish at high energies. \newline 
In the correspondent smooth system the potential,$V=V_1(q_1)+V_2(q_2)$  is separable. Thus, its wavefunctions, $\Psi^{sm}_n$, can be written as a product of the wavefunctions of $H_i$: $\{\Psi^{sm}_n\}_{n=1}^\infty=\Psi_{1,k_1}(q_1)\Psi_{2,k_2}(q_2)$ where  $\{\Psi_{i,k_i}\}_{k_i=1}^\infty$ are the wavefunctions of the smooth one dimensional Hamiltonian $H_i$ and $E^{sm}_{n(k_1,k_2)}=E_{k_1}+E_{k_2}$. \newline Since $V_{i}$ are even:
\begin{align}
\Psi_{i,k_i}(q_{i})= \begin{cases} \Psi_{i,k_i}(-q_{i}) \phantom{-} \text{if $k_i$ is even} \\ -\Psi_{i,k_i}(-q_{i}) \phantom{-} \text{if $k_i$ is odd}
\end{cases}
\end{align}
When both $k_1$ and $k_2$ are odd, the series of wavefunctions $\{\Psi^{sm}_{n_j(k_1,k_2)}\}_{n_j=1}^\infty$ vanishes on both axes, hence, the non-smooth Hamiltonian for the case of step at the origin has a subsequence of wavefunctions of the form: 
% $n=1,2...$ odd functions step $1=n_1(1,1),5=n_2(1,2),9$ odd function of smooth $1,7=\tilde n_2(1,2),11$
\begin{align}
    \Psi^{S_0}_{ n_j(k_1,k_2)}(q_1,q_2)=
    \begin{cases}
    \begin{split}
    \Psi^{sm}_{\tilde n_j(k_1,k_2)}=&\Psi_{1,k_1}(q_1)\Psi_{2,k_2}(q_2),\\ &(q_1,q_2)\in \mathbb{R}^{2}/S_0      \end{split} \\
    \\ \label{product}
    \phantom{-} \phantom{-} 0 \phantom{-} & (q_1,q_2)\in S_0 
    \end{cases}
\end{align} 
These solutions are smooth in the domain ($\mathbb{R}^{2}/S_0$) and satisfy Dirichlet boundary conditions on $S_0$. Moreover, $\Psi^{S_0}_{ n_j(k_1,k_2)}$ concentrates on the projection of classical level sets; as the one-dimensional wavefunctions are well approximated by the WKB approximation \cite{brack2018semiclassical}, they decay exponentially outside of the classical allowed region of motion:
\begin{align}
\label{wkb}
\displaystyle \Psi_{i,k_i} (q_1)\approx C_{0}{\frac {e^{\theta +i \hbar ^{-1}\int {\sqrt {2\left(E_{i,k_i}-V_i(q_i)\right)}}\,dq_i}}{\hbar ^{-1/2}{\sqrt[{4}]{2\left(E_{i,k_i}-V_i(q_i))\right)}}}}.
\end{align}
Next we show that the fraction of such odd wavefunctions for the case of a step at the origin is $1/3$.
From equation \ref{eqn:ebk} for the smooth case (i.e. $b=0$) we deduce that   wavefunction that are odd in both directions (odd $k_1,k_2$) constitute one quarter of all wavefunctions:
\begin{align}
 \lim_{E\to\infty} \frac{\#\{\Psi^{sm}_{\tilde n_j(k_1,k_2)}: E_{\tilde n_j}=E^{1}_{k_1}+ E^{2}_{k_2} \leq E \}}{\#\{\Psi^{sm}_n:E_{n} \leq E\}}=\frac{1}{4}.
\end{align}
Since the step is at the origin:
\begin{align}
\text{Vol}(E)^{S_0}=\frac{3}{4}\text{Vol}(E)^{sm}
\end{align}
and thus, by Weyl's law
\begin{align}
\lim_{E\to\infty} \frac{\#\{\Psi^{S_0}_{n_j}: E_{n_j} \leq E \}}{\#\{\Psi^{S_0}_n:E_{n} \leq E\}} =
\lim_{E\to\infty} \frac{\#\{\Psi^{sm}_{\tilde n_j}: E_{\tilde n_j} \leq E \}}{\frac{3}{4}\#\{\Psi^{sm}_n:E_{n} \leq E\}} =
\frac{1}{3}.
\label{fraction}
\end{align}
We conclude that for a step at the origin there is no quantum ergodicity in configuration space, and, in fact, there is a positive measure set of eigenfunctions that concentrate on the classical level sets. 

To examine the behavior for non-symmetric pseudointegrable cases, we study numerically the shifted corner in the harmonic case: we find the level spacing of the eigenvalues and study the projections to configuration space of the eigenfunctions. Both studies propose that the shift does not break the concentration of a large subset of eigenfunctions on classical level sets.

\begin{figure}[htbp]
 \centering
 \begin{subfigure}[t]{.4\textwidth}
  % include second image
  \includegraphics[width=1\linewidth]{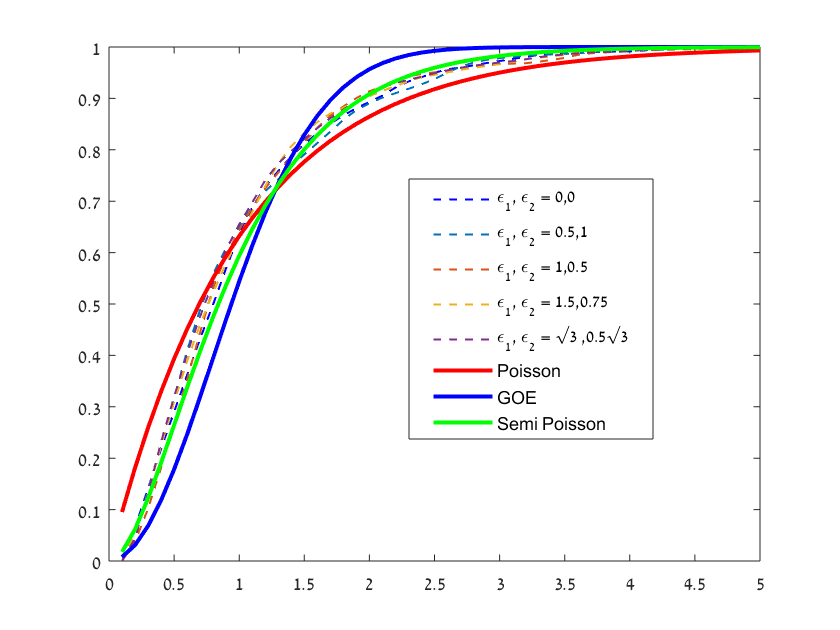}  
  \caption{} 
  \label{fig:cdf}
\end{subfigure}
\begin{subfigure}[t]{.4\textwidth}
  % include second image
  \includegraphics[width=1\linewidth]{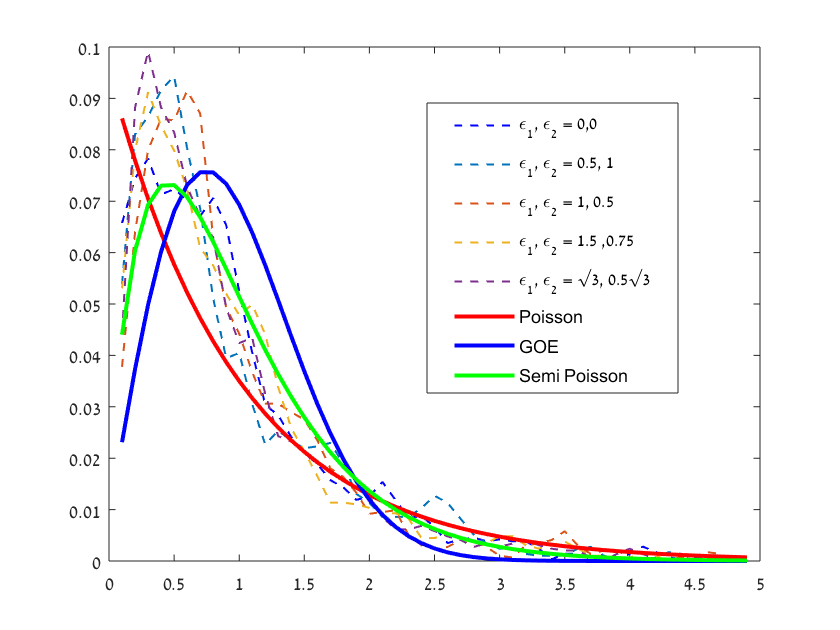} 
    \caption{}
      \label{fig:pdf}
      \end{subfigure}
      \caption{PDF and CDF of the level  spacing for a non-resonant Hamiltonian for several positions of the step. The semi-Poisson distribution (solid thick green line) provides the best fit for all positions of the step (dashed lines), including a step at the origin (blue dashed line). (a) Cumulative distribution functions of Poisson, semi-poisson GOE and numerically calculated CDFs  (b) Probability density functions of Poisson, semi-poisson GOE and numerically calculated PDFs.  The level spacing are found by a finite differences scheme for the time independent Schrodinger Eq. for the Hamiltonian \ref{eq:modelsham} with $V=U_0+U_1$ where $U_0=\frac{q_1^{2}}{2}+\frac{\omega_2^{2} q_2^{2}}{2}$ and $U_1=-\epsilon_{1}q_1-\epsilon_{2}q_2$. The step is located at the origin and is numerically represented as $V=10^{28}$. Here,  $\omega_1=1,\omega_2=\sqrt{2}$ and  $(\epsilon_1, \epsilon_2)=(0,0),(0.5,0.25),(1,0.5),(1.5,0.75),(\sqrt{3},\frac{\sqrt{3}}{2})$.}
      \label{disp}
      \end{figure}

 It is convenient for the study of the non-symmetric system to keep the step at the origin and shift the original harmonic potential to have a minimum at $(\epsilon_1,\frac{\epsilon_2}{\omega_2^{2}})=\epsilon \cdot (\cos \alpha,sin\alpha)$. Then the potential is of the form: $V=U_0+U_1$ where $U_0=\frac{q_1^{2}}{2}+\frac{\omega_2^{2} q_2^{2}}{2}$ and $U_1=-\epsilon_{1}q_1-\epsilon_{2}q_2$. Here, $\epsilon=0$ corresponds to the system with a step at the origin, and we study the behavior  for a non-resonant case at finite values of $\epsilon$, beyond the small perturbation regime. 
 Figure \ref{fig:cdf}  compares the cumulative mean level spacing distribution of this shifted potential of the first 1500 energy levels to the cumulative Poisson distribution (characterizing integrable systems, $N_p(s)=1-e^{-s}$, reflecting their locality in the classical phase space) and to the cumulative random matrix ensembles distribution, GOE (characterizing chaotic systems,  $N_W(s)=1-e^{-\frac{\pi s^2}{4}}$, reflecting their non-local nature in the classical phase space). We obtain intermediate statistics as in pseudo integrable billiards, close to semi-Poisson distribution ($N_{sp}(s)=1-e^{-2s}(2s+1)$) \cite{bogomolny1999models} (such a behaviour was also observed in a certain range of parameters in step-like time dependent one d.o.f. Hamiltonian \cite{garcia2006semi}). 
 
 Figure \ref{fig:cdf} shows that the dependence of the level spacing on $\epsilon$ appears to be mild and similar to the case $\epsilon=0$. Recall that in the case of a step at the origin, we showed that there is a positive density sequence of eigenfunctions concentrated on classical level sets. Namely, the level spacing distribution at $\epsilon=0$ reflects this locality in phase space, together with the non-locality associated with pseudointegrability.  Fig. \ref{fig:cdf} suggests that this behaviour persists  when the step is shifted from the origin.
 In fact, Fig. \ref{fig:pdf} shows that the distribution with the largest repulsion is achieved at $\epsilon=0$.   
      
\begin{figure}
 \centering
 \begin{subfigure}[t]{.4\textwidth}
  % include second image
  \includegraphics[width=1\linewidth]{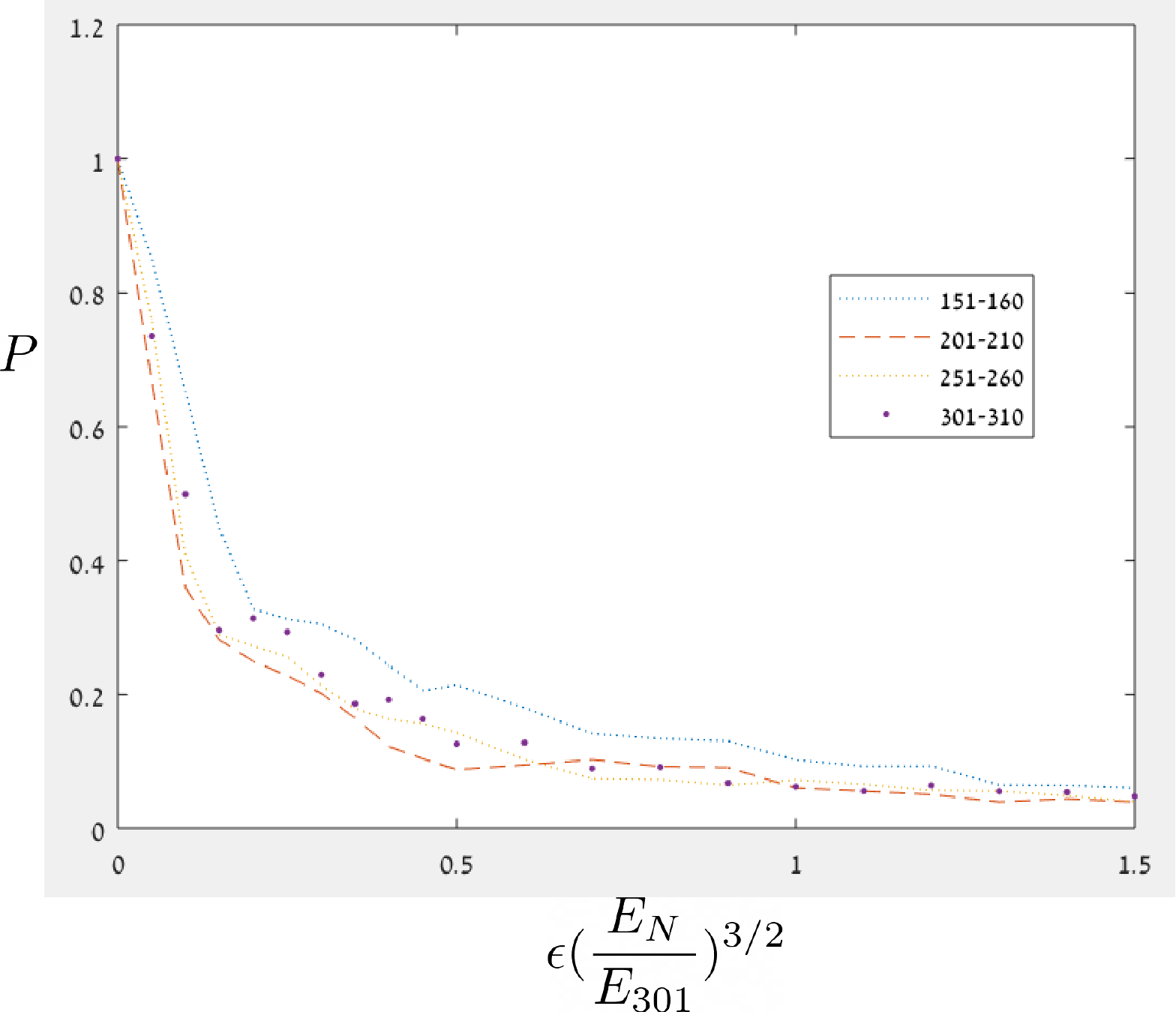}  
  \caption{}
  \label{maincomponent}
\end{subfigure}
\begin{subfigure}[t]{.4\textwidth}
  % include second image
  \includegraphics[width=1\linewidth]{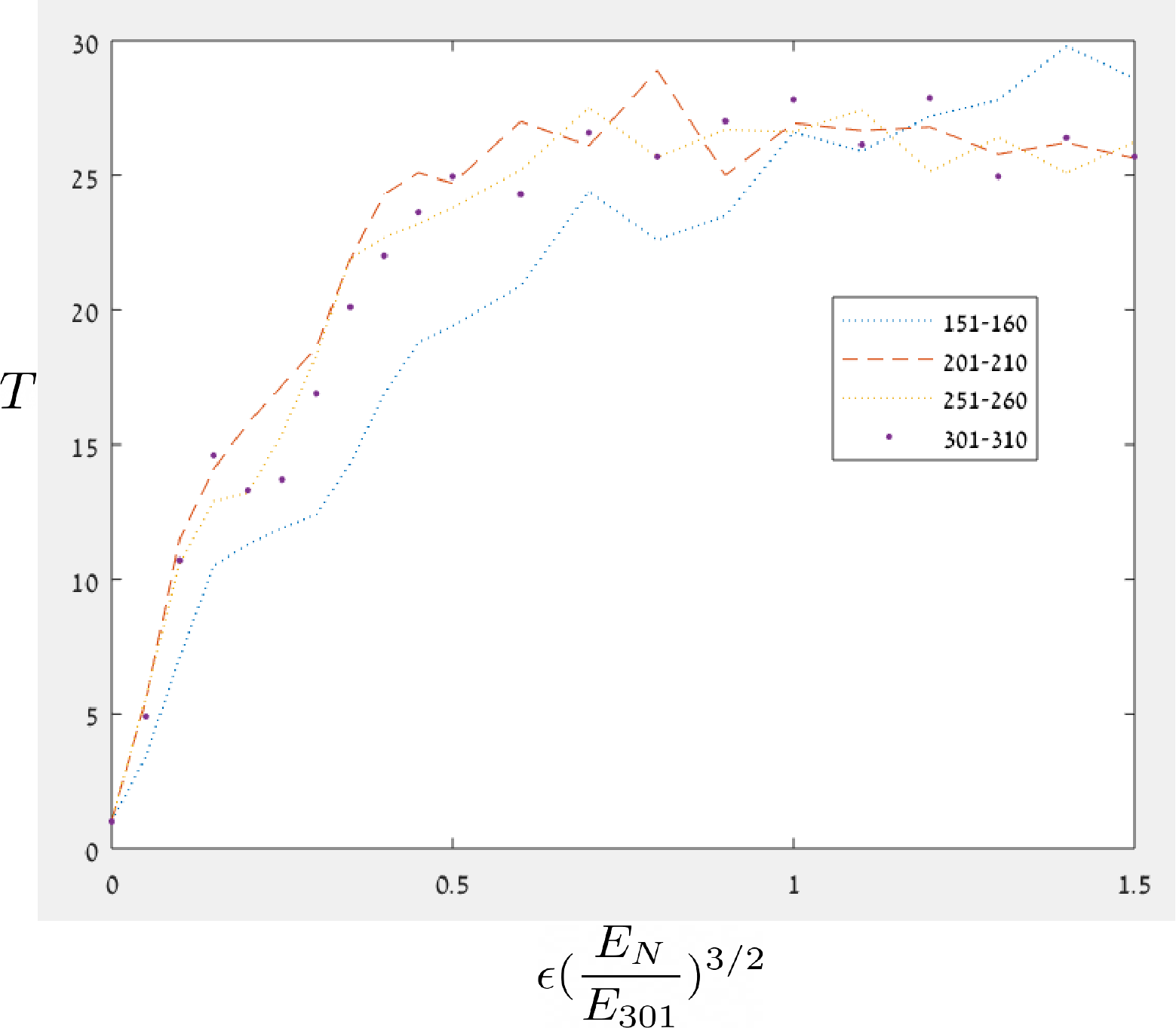}
  \caption{}
  \label{mixingrate}
\end{subfigure}
    \caption{Scaling of the perturbed wavefunctions with $\epsilon$ and energy. (a) The  mean maximal projection on unperturbed wavefunctions along the energy-scaled $\epsilon$, $\epsilon (\frac{E_N}{E_{301}})^{3/2}$:  $P (\epsilon (\frac{E_N}{E_{301}})^{3/2},N;10,400)$ (b) The mean number of above-threshold contributing unperturbed wavefunctions along the energy-scaled $\epsilon$, $\epsilon (\frac{E_N}{E_{301}})^{3/2}$:  $T (\epsilon (\frac{E_N}{E_{301}})^{3/2},N;10,400,0.01)$. These functions are plotted  
    for $N=151,201,251,301$ for and for several $\epsilon=(\epsilon_1,\epsilon_2=\frac{\epsilon_1}{2})$ values.    }
      \label{mixing}
      \end{figure}

To substantiate the claim that, as suggested by the level spacing plots, at large energies, the general step system still has a positive fraction of wavefunctions that concentrate on classical level sets, we calculate the wavefunctions for such systems. Since the wavefunctions depend continuously on $\epsilon$, for any given maximal  energy, for small enough $\epsilon$, such a fraction of concentrated wavefunctions exists. Hence, we first find the natural scaling of $\epsilon$ with $E$ and establish that our wavefunction calculations are far from the trivial limit of ${\epsilon \to 0}$, namely, that the perturbed wavefunctions do not correlate well with unperturbed wavefunctions.  \newline Expanding the wavefunctions in $\epsilon$,  the first order correction to $|n(\epsilon)\rangle=|n^{(0)}\rangle+ \epsilon| n^{(1)}\rangle+ O(\epsilon^2)$, is: \newline
${\displaystyle \epsilon |n^{(1)}\rangle =\sum _{k\neq n}{\frac {\langle k^{(0)}|U_1|n^{(0)}\rangle }{E_{n}^{(0)}-E_{k}^{(0)}}}|k^{(0)}\rangle } \newline
$where$ \phantom{-} U_1=-\epsilon_1 q_1-\epsilon_2 q_2$. So for large energies, the number and power of terms that contribute significantly to the sum are expected to stabilize provided we use the scaling:
 $\epsilon_1\propto\frac{E_{n+1}-E_{n}}{q_1}$ and $\epsilon_2\propto\frac{E_{n+1}-E_{n}}{q_2}$. 
Since,  for harmonic oscillators, $q_i \propto\sqrt{E}$ and  $N(E)\propto \text{Vol}(E)\propto E^{2}$, so $E_{n+1}-E_{n}\propto\frac{1}{E}$, 
we conclude that the stabilization is achieved  provided 
 $\epsilon\propto\frac{1}{E^{1.5}}$. As higher orders of the perturbation series give the same result, we actually expect that  
  ${\displaystyle |n(\epsilon)}\rangle-{\displaystyle |n^{(0)}\rangle} = O(\epsilon E_{n}^{1.5})$. 
To capture the distance between eigenfunctions of the non-perturbed Hamiltonian to the perturbed one around an energy level $E_N$, we calculate $P$, the mean squared maximal projection on unperturbed wavefunctions, and $T$, the mean number of above-threshold contributing unperturbed wavefunctions: 
\begin{equation}\label{mix}
\begin{split}
P (\epsilon,N;\Delta N,J) &=\frac{1}{\Delta N}\sum_{n=N}^{N+\Delta N} \max_{j^{0}\le J}{{|\langle  \displaystyle j^{0} \displaystyle |n(\epsilon)}\rangle}|^{2}\\
T(\epsilon,N;\Delta N,J,\delta)&=\frac{\sum_{n=N}^{N+\Delta N}  \# ({{|\langle  \displaystyle j^{0} \displaystyle |n(\epsilon)}\rangle|}^{2}>\delta)}{\sum_{n=N}^{N+\Delta N} \sum_{j^{0}=0}^J {\langle  \displaystyle j^{0}\displaystyle |n(\epsilon)}\rangle^{2}}.
\end{split}    
\end{equation}
 \newline
 Figure \ref{mixing} shows that 
 $P (\epsilon E_N^{3/2},N;\Delta N,J)$ and $T (\epsilon E_N^{3/2},N;\Delta N,J,\delta)$ are, to a good approximation,  independent of $N$,
 supporting the validity of our scaling. 
Moreover, while for small  $\epsilon (\frac{E_N}{E_{301}})^{3/2}$ we see that, as expected, there is a strong correlation between the perturbed and unperturbed wavefunctions,   for 
$\epsilon E_N^{3/2} \ge  E_{301}^{3/2} $ the maximal projection, $P$, is small while the level of mixing, $T$, is large, indicating that for such values of $\epsilon E^{3/2}$ we are indeed far from the small $\epsilon$ limit.   Additional computations show that a further increase in $\epsilon E_N^{3/2}$ leads to further decrease in $P$. 
      
\begin{figure}
 \centering
 \begin{subfigure}[t]{.6\textwidth}
  % include second image
  \includegraphics[width=0.8\linewidth]{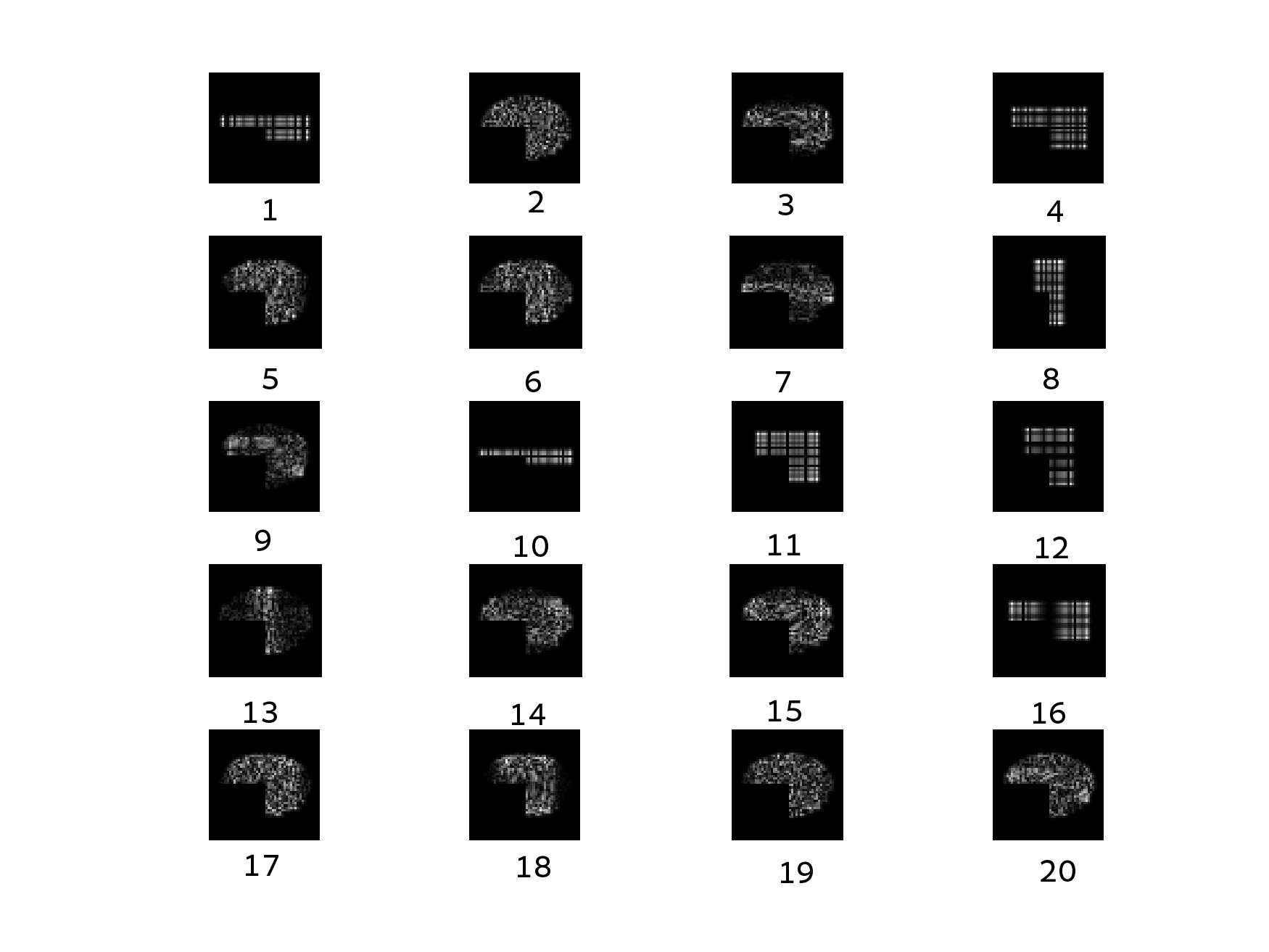}  
  \caption{ }
  \label{mediumpert}
\end{subfigure}
\begin{subfigure}[t]{.6\textwidth}
  % include second image
  \includegraphics[width=0.8\linewidth]{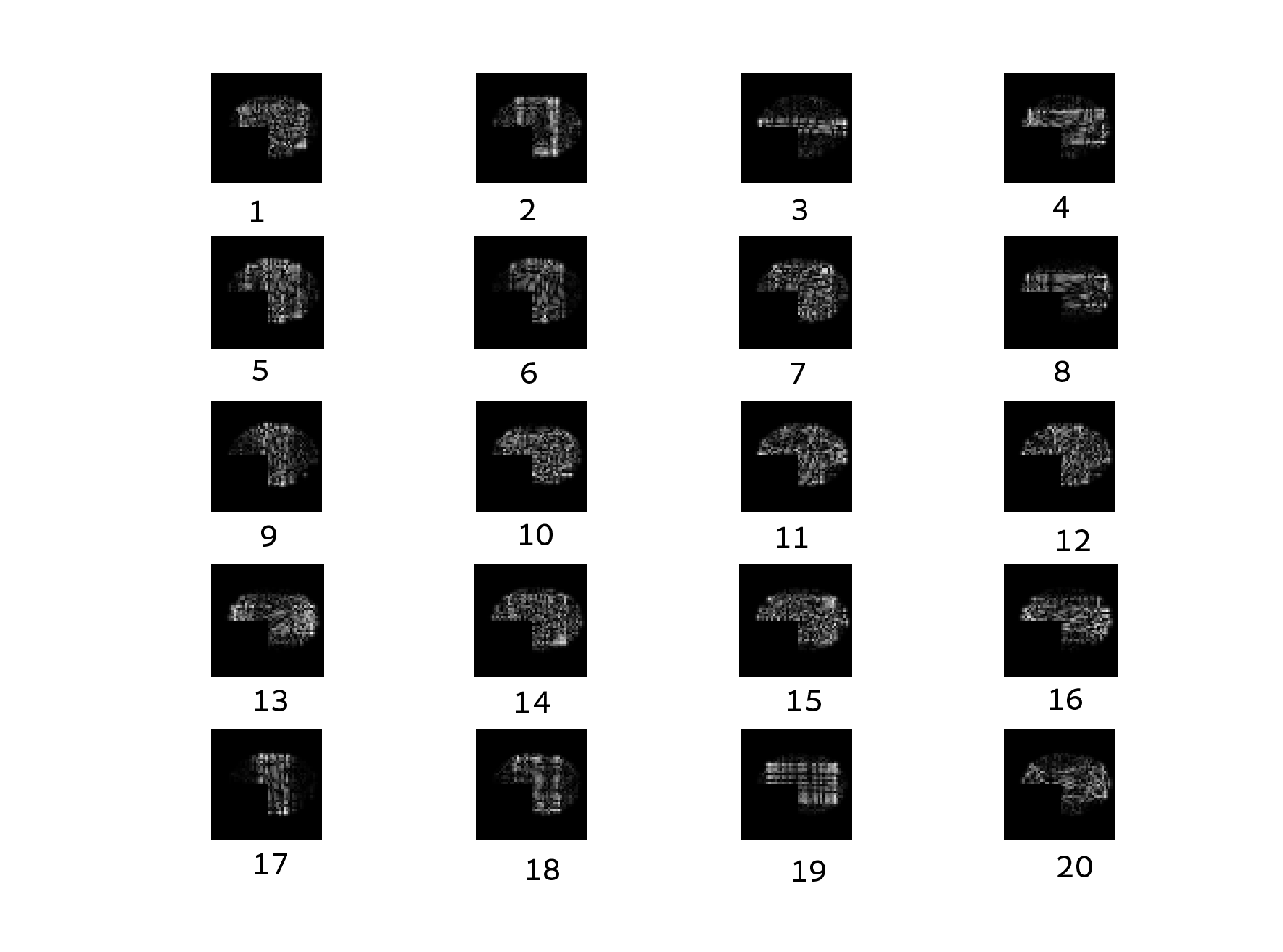}  
  \caption{ }
  \label{bigpert}
\end{subfigure}
    \caption{High energy wavefunctions for a step at the origin and for a shifted step.  (a) The unperturbed Hamiltonian.   (b) The perturb Hamiltonian  with $(\epsilon_1,\epsilon_2)=(1.5,0.75)$. The wavefunctions  for $n=1481-1500$ are plotted. To better visualize the main mass concentration we plot   
    $Log(|\Psi_n(q_1,q_2;\epsilon)|+\max_{q_1,q_2}|\Psi_n((q_1,q_2;\epsilon)|)$. }
      \label{wavefunctions}
\end{figure}

Finally, we show that even when   $\epsilon (\frac{E_{n}}{E_{301}})^{3/2}\gg 1$, i.e. when the wavefunctions are not well approximated by the unperturbed wavefunctions,
a substantial fraction of the  wavefunctions concentrate on classical level sets.
Figure \ref{wavefunctions} shows  the 1481-1500 wavefunctions in Logarithmic scale normalized by the maximal absolute value of the wavefunctions for the unperturbed (step at the origin) and perturbed ($\epsilon=(1.5,0.75)$) wavefuncations (so $\epsilon (\frac{E_{1500}}{E_{301}})^{3/2}=5.25$). For both the perturbed and unperturbed systems,  wavefunctions that are concentrated along the  classical level sets, i.e., are essentially restricted to the configuration space region $(q_1,q_2) \in [q_1^{min}(E_1,\epsilon_1),q_1^{max}(E_1,\epsilon_1)]\times [q_2^{min}(E_2,\epsilon_2,\omega_2),q_2^{max}(E_2,\epsilon_2,\omega_2)]
\setminus S_{q^{wall}}$ where $q_i^{max,min}$ correspond to the classical level set boundaries,  are clearly seen (e.g. see  wavefunction 1 in the unperturbed system and wavefunction 19 in the pertubed system). We call such wavefunctions concentrated wavefunctions.

To quantify this observation, we need to distinguish between concentrated wavefunctions from wavefunctions which are not concentrated. To this aim we define vertical and horizontal means of the wavefunctions: 
\begin{equation}
\begin{split}
M^H_n(q_2)&=\int_{-\infty}^{\infty}|\Psi_n(q_1,q_2)|^{2}dq_1 \\
M^V_n(q_1)&=\int_{-\infty}^{\infty}|\Psi_n(q_1,q_2)|^{2}dq_2.     
\end{split}    
\label{means}
\end{equation}
and suggest that
\begin{equation}
\label{normalizedenergy}
\Tilde{E}=\frac{V_1(arg \max_{q_1}M^V_n(q_1))+V_2(arg \max_{q_2}M^H_n(q_2)}{E}
\end{equation}
provides a good indicator for the wavefunctions concentration: it is close to one for concentrated wavefunctions and has a much lower value for the rest of the wavefunctions.

\begin{figure}
\centering
\includegraphics[width=0.9\linewidth]{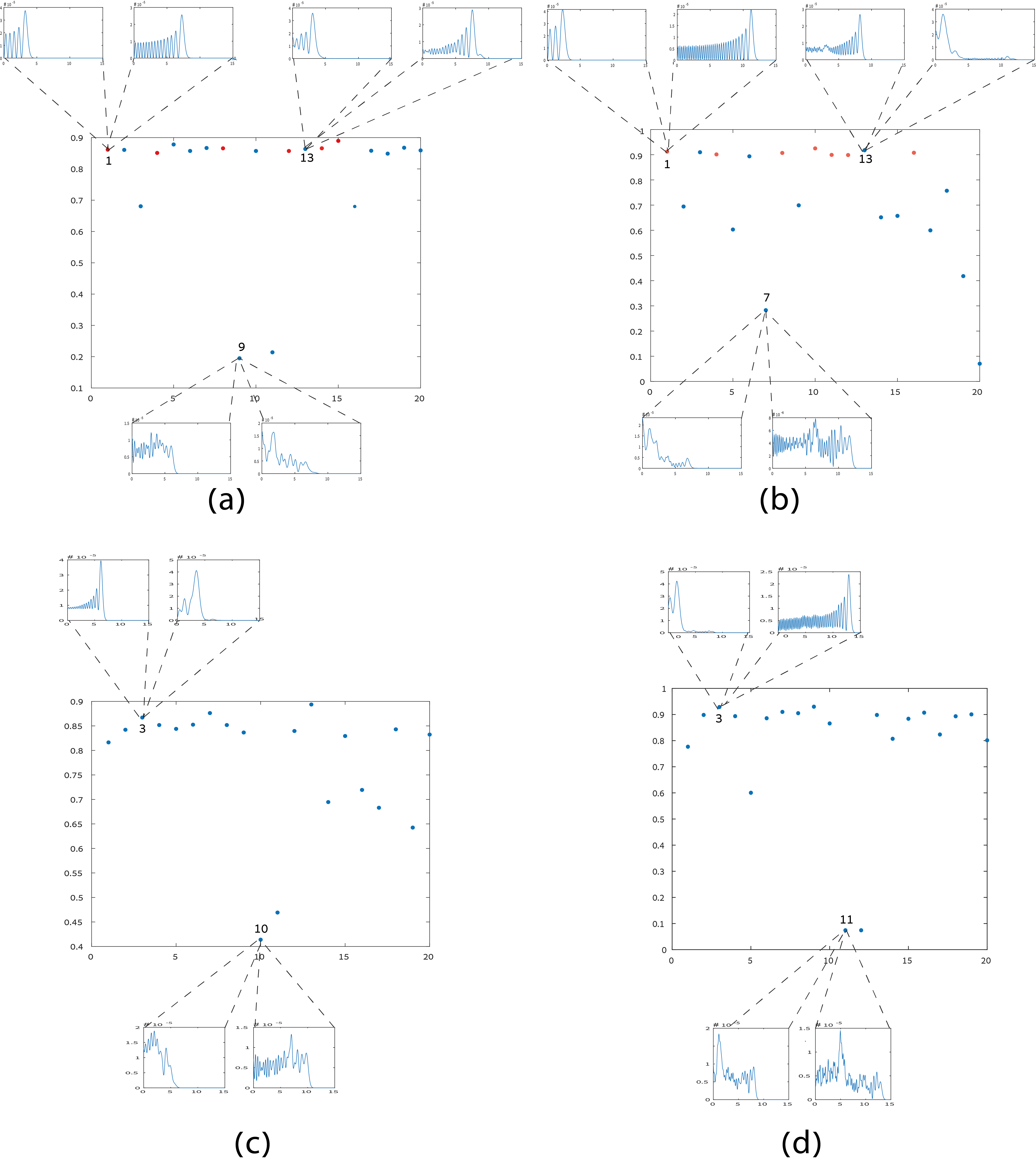}
    \caption{An indicator for the concentration of wavefunctions on classical level sets. The indicator $\Tilde{E}$ of Eq. \ref{normalizedenergy} is plotted for the case of a step at the origin, (a) 481-500 and  (b) 1481-1500 wavefunctions (the wavefunctions of Fig. \ref{wavefunctions}a). The indicator $\Tilde{E}$ is plotted for the case of a shifted step ( $\epsilon=(1.5,0.75)$)  
   (c) 481-500 and (d) 1481-1500 wavefunctions(the wavefunctions of Fig. \ref{wavefunctions}b). \newline
   The insets present $M^H,M^V$ for specific points}
      \label{etilde}
      \end{figure}

Figures \ref{etilde}(a,b)  present  $\Tilde{E}$ values in the case of corner at the origin for low (a) and high (b) ranges of energies. Red points represent $\Tilde{E}$ values for the product wavefunctions of Eq. (\ref{product}) and constitute around $1/3$ of the  20  $\Tilde{E}$ values.
We see that some of the blue points align with the red ones, while others, around 1/5 for the lower energies and 1/2 for the higher energies have a much lower value. The insets present $(M^H_n,M^V_n)$ in the positive quadrant for the three different types of wavefunctions: for a product  wavefunction (red point, wavefunction 1 in \ref{etilde}(a) ), for a concentrated wavefunction with a similar $\Tilde{E}$ value (blue point, wavefunction 13 in \ref{etilde}(a) ) and for a non-concentrated wavefunction with a low $\Tilde{E}$ value (blue point, wavefunction 9 in \ref{etilde}(a) ). 
In the first two cases we recognize an oscillatory structure within the classically allowed region, and we observe that the maximal power appears close to the edge. In contrast,  the insets corresponding to the low $\Tilde{E}$ value show a non oscillatory structure with peaks at arbitrary positions within the Hill region. \newline
Figures \ref{etilde}(c,d) present a similar computation for the case of the shifted potential,    $\epsilon=(1.5,0.75)$, for which there are no product wavefunctions, yet concentrated and not concentrated wavefunction do appear, and the indicator  $\Tilde{E}$ seems to distinguish between these two types of wavefunctions. 

The reasoning for this suggestion is as follows;
For step at the origin, for the product wavefunctions (eq. \ref{product}), $M^H_n(q_2)=|\Psi_{n,2}(q_2)|^2$ for $q_2>0$ and $M^H_n(q_2)=|\Psi_{n,2}(q_2)|^2/2$ for   $q_2<0$, so by the WKB approximation (eq.\ref{wkb}), and similarly for $M^V_n(q_1)$,  we indeed expect $\Tilde{E}=1-f(E)$ for some function $f(E)$ which tends to zero as $E$ goes to infinity (e.g., Figures \ref{etilde}(a,b) suggest that $f(E_{500})\approx0.15, E_{500}=39.9 $ and $f(E_{1500})\approx0.1, E_{1500}=70.5$). 
For non-product yet concentrated wavefunctions on some classical configuration space region defined by the partial energies $(E_1,E_2)$,
%we get $\Psi_n(q_1,q_2)\approx \tilde \Psi_n(q_1,q_2)w_n(q_1,E_1)w_n(q_2,E_2)$ 
the $arg max$ of  $M^{V,H}$ cannot be larger than the corresponding $q_i^{max}$. Moreover, as classically, one of the momenta components vanishes at the edges of the classical region, the projection of the Liuoville measure to the configuration space there is expected to be larger, hence, by the correspondence principle, we expect maximal densities near the edges. Hence,  $\Tilde{E}$ provides the approximate ratio between the sum of the potential energies at the classical region corners (belonging to the boundary of the classical Hill region) to the total energy,  so we expect it to have a similar $\Tilde{E}$ values to the corresponding product wavefunctions.
 In contrast, for a wavefunction which does not concentrate on a single classical level set we do not expect the maxima in the horizontal and vertical directions to lie necessarily on the boundary of the Hill region (see insets corresponding to the lower $\Tilde{E}$ values), thus the sum of the potential energies at such an interior point leads to a lower value of $\Tilde{E}$.

In conclusion, Figures \ref{wavefunctions}  and \ref{etilde} suggest that the fraction of concentrated wavefunctions does not vanish at high energies even when the step is shifted. \newline

 %summary     
     Summarizing, we studied the correspondence of a quantum step-oscillator - a two dimensional quantum oscillator in the presence of a step (a step-like region $S$ in the configuration space  at which the potential energy is infinite) to its classical analog,  a pseudointegrable Hamiltonian impact system.   For the case of harmonic resonant oscillators with a corner at the origin, for which families of periodic orbits can be explicitly constructed, we demonstrated that the EBK quantization condition provides a good predictor to the energy levels (Figure \ref{figtable}), and that Weyl's law provides a good approximation to the growth in the number of wavefunctions (Figure \ref{fig2}). Moreover, we observed that in even-resonance cases two different families of periodic orbits belonging to the same component of the level set co-exist, with distinct corresponding wavefunctions, each contributing a positive portion to the phase space volume (Figure \ref{fig2}).   This demonstrates that the non-ergodicity of level sets has a quantum analog. We showed that the intermediate level spacing of the quantum step-oscillator for non-resonant and not necessarily harmonic potential hardly depends on the position of the step (taken in the negative quadrant) and is approximately  semi-Poisson, indicating repulsion of energy levels, similar to the level spacing obtained for pseudointegrable billiards (Figure \ref{disp}).
     When the step is at the origin, we showed that there is a positive fraction of wavefunctions that remain concentrated along the classical level sets at arbitrarily high energies, as occurs for integrable systems, namely they do not tend to equidistribute  in the configuration space as is the case for pseudointegrable billiards (Eq. (\ref{product})-(\ref{fraction}) and Figures \ref{wavefunctions}a and \ref{etilde}a,b). Finally, when the corner is shifted from the origin, we conjecture, based on numerical evidence for non-resonant harmonic oscillators,  that there is a positive density series of wavefunctions which are not equidistributed and concentrated along the classical level sets (Figures \ref{wavefunctions}b and \ref{etilde}c,d).

%conj:
%For a general quantum step-oscillator, it is expected that classical resonant %level sets, which here, may contain both periodic and a-periodic components, will, %by the EBK condition, also  have  concentrated wavefunctions along these families, %with two types of such functions for some of the resonant surfaces. As finding %such resonant surfaces for the general case is challenging, we used perturbation %theory to examine the behavior for general systems at which the step is shifted %from the origin. 

%why should you care
Classical Hamiltonian systems with impacts model systems in which strong short range repulsions (such as atomic forces) are combined with attracting forces (such as Van der Waals forces)  \cite{lerman2012saddle}.  Such systems are integrable when the repulsion and attracting forces have sufficiently many common symmetries, and can become pseudointegrable when such symmetries occur along surfaces with corners \cite{becker2020impact,pnueli2021structure}.  Here we propose that the  correspondence between such quantum systems and their  classical  analogs can be studied using both integrable quantization methods (EBK and WKB) and methods used in the study of pseudointegrable billiards (level spacing).  The implications of these observations on  quantum system that arise in applications, and, in particular, the asymptotic dependence on parameters governing the impact surface geometry (i.e. the singular limit by which corners become smooth), the Erenfest time and the evolution of wave packets for such systems is challenging and is left for future studies. The quantum step-oscillators system provides a rich yet simple toy model for studying such questions.

% We show that our systems have intermediate level-spacing statistics similar to polygonal billiards. Yet, we prove that unlike the billiards case, superscars structure in configuration space doesn't disappear in large energies at least for some limit cases. We observe numerically that it is likely to have the same superscars structure for the general case as well. 

\nocite{*}

Acknowledgments: VRK is the Estrin family chair of computer science and applied mathematics. We thank the support of ISF grant 787/22. We also thank M. Aizenman, D. Mangoubi,  and U. Smilansky for stimulating discussions.

%\clearpage

\bibliography{apssamp}% Produces the bibliography via BibTeX.

\end{document}